\documentclass[journal]{IEEEtran}
\usepackage{graphicx}
\usepackage{color}

\definecolor{dkgreen}{rgb}{0,0.6,0}
\definecolor{gray}{rgb}{0.5,0.5,0.5}
\definecolor{mauve}{rgb}{0.58,0,0.82}

\newcommand{\gclass}[1]{\texttt{#1}}
\begin{document}

\title{Extending Geant4 Parallelism with External Libraries (MPI, TBB) and Its Use on HPC Resources}

\author{Andrea~Dotti,
Makoto~Asai,
Guy~Barrand,
Ivana~Hrivnacova,
Koichi~Murakami,
\thanks{A.~Dotti and M.~Asai are with SLAC National Accelerator Laboratory, Menlo Park, CA 94025 (USA) (corresponding author~-~telephone: 650-296-2866, e-mail adotti@slac.stanford.edu). }%
\thanks{G.~Barrand is with LAL Orsay, 91809 Orsay (France).}%
\thanks{I.~Hrivnacova is with IN2P3/IPN Orsay, 91406 Orsay (France).}%
\thanks{K.~Murakami is with KEK, Tsukuba, Ibaraki 305-0801 (Japan).}
}

\maketitle
\pagestyle{empty}
\thispagestyle{empty}


\section{Introduction}
\IEEEPARstart{T}{he} emergence of multi- and many-core processors has been a well-established
trend in the chip-making industry during the past decade.  While this paradigm
guarantees the continued increase of CPU performance, it requires some
adaptation of existing code in order to better utilize these architectures.\\
Geant4 is a toolkit for Monte Carlo simulation of the transportation and interaction of particles in matter~\cite{Agostinelli2003250}.
It can be used in a wide variety of applications including high energy physics, space and medical science.\\
With the release of Geant4 Version 10.0 in December 2013 event-level parallelism has been introduced~\cite{1742-6596-513-2-022005}.\\
A Geant4 application is defined by the use of an instance of the
\gclass{G4RunManager} class or of a user defined class derived from it.  This
class defines the main interaction with the user: it provides interfaces to
define the {\it user initializations} (e.g. geometry and physics definitions)
and the {\it user actions} that permit interaction with the simulation kernel
and retrieve output information.  In particular, \gclass{G4RunManager} provides
the interface to start the simulation of a run, which is a collection of events.
For multithreaded applications a derived class \gclass{G4MTRunManager} is used
allowing the number of worker threads to be specified. When a new run is requested it is
the responsibility of \gclass{G4MTRunManager} to start and configure worker
threads.  Each thread owns an instance of \gclass{G4WorkerRunManager} and
shares only the user initialization classes, while it owns a private copy of the
user action classes.  Workers continue to request events from the master until
there are no more events left in the current run.  At the end of the run the
results collected by threads can be merged in the global run.\\
To maximize the compatibility with different systems multi-threading in Geant4 is implemented via the POSIX {\it pthread} library primitives. In this paper we present extensions to Geant4 that allow for integration with additional well established parallelization frameworks.

\section{TBB Integration}
The user can overwrite the default implementation of the threading model in Geant4 providing a user-derived \gclass{G4RunManager} (and other helper classes). This approach is demonstrated within the  Gent4 TBB example, that replaces {\it pthreads}.\\
The Intel TBB library allows for the development of a task-based parallelism. When using TBB, a user should provide a user class derived from the base class \gclass{tbb::task} and implement a virtual method \gclass{execute}.\\
One of the characteristics of TBB is the mapping of tasks to the execution threads. It is done by the library dynamically at run-time and the user has no control over the underlying threads from a {\it task} object. This design poses some challenges to run a Geant4 job based on TBB. In our case threads need to be initialized and the thread-local-storage properly configured to comply with the Geant4 master/worker model~\cite{Ahn2013}.\\
Few new classes 
, specific to TBB, have been introduced in Geant4 (to be released in Geant4 Version 10.2 under {\it examples/extended/parallel/TBB} directory).\\
With the latest version of TBB it is possible to control thread initialization via the class \gclass{tbb::task\_scheduler\_observer}, we use this feature to initialize specific thread-local-storage. The Geant4 master needs to live in its own thread, to accomplish this \gclass{tbb::task\_arena} instances can be used to partition the thread pool into two and assign a single thread to the master. The role of the master now is to create a list of \gclass{tbb::task}s that are passed to TBB run-time. Each task, when executed, will retrieve the thread-specific {\it worker-run-manager} and will accomplish the simulation of one or more events. This design has allowed the CMS Experiment at CERN to migrate the simulation framework (based on Geant4) to a parallel, TBB based, one. The ATLAS experiment at CERN is also investigating the possibility to use TBB as the parallelization library in their experiment software framework.\\
While TBB does not bring by itself any particular benefit in terms of CPU or functionality, this development will allow for an easier integration of Geant4 into large software projects where a task-based parallelism is adopted. 

\section{MPI Interface}
Support for MPI parallelism is available in Geant4 since some time ({\it examples/extended/parallel/MPI} directory), recently we have extended this example to provide hybrid applications that use both MPI and MT. To efficiently reduce the memory consumption it is possible to schedule a single MPI job for each node and use Geant4 multi-threading capabilities to scale across the CPUs and cores available on the node.\\
To activate MPI in Geant4 it is enough to create instances of  the two classes \gclass{G4MPImanager} and \gclass{G4MPIsession}. The user-interaction is performed via the usual Geant4 UI commands: a {\it /run/beamOn} command will trigger the MPI ranks to cooperatively perform the simulation of the run.\\
Up to Geant4 version 10.1 (December 2014), the use of MPI in a job is limited to the steering of the job: the {\it rank 0} accepts UI commands from the user, distributes work among all ranks and controls the random number generator seeds. Since Geant4 Version 10.0 a new module is available in Geant4 to support user-analysis~\cite{g4ana}: histograms and ntuples can be created and saved to different file formats (ROOT, AIDA XML and CSV). A characteristics of the g4analsysis package is its thread-safety and its ability to automatically  reduce histograms at the end of the job: to minimize the use of mutex/locks each thread has its own copy of the analysis histograms, at the end of the job these are automatically reduced in the master thread and written to a single file. With the upcoming Geant4 Version 10.2 histograms in the g4analysis package will support streaming via MPI.\\
To simplify the user handling of histograms the following strategy is implemented:   
\begin{itemize}
\item[1] Each thread of each rank owns its copy of the analysis histograms. These are filled independently during the event loop and no synchronization between threads or ranks is required
\item[2] At the end of the run, all working threads belonging to the same rank reduce the histograms in the master thread
\item[3] The MPI ranks with id $>0$ merge the histograms with into the rank with id 0 using MPI messages
\item[4] The MPI rank with id=0 is responsible of writing the final output file to disk
\end{itemize}
We are investigating improved MPI communication patterns to reduce the time needed at the end of the job to merge histograms. The improved code will be released with Geant4 Version 10.2 (scheduled for December 2015).

\section{Use of HPC resources}
Traditionally HPC resources, consisting of a very large number of nodes and processing units, are suited to MPI applications. In some cases MPI is the preferred way to run parallel jobs on such systems.\\
Up to Version 10.0 the use of Geant4 on these systems has been limited by its lack of parallelism: the only way to scale among the cluster was to use a multi-process approach in which for each computing unit a clone of the process was started. It was left to the user to create scripts to achieve this. In some cases, when the memory used by each process is large (e.g. LHC experiments), it was not possible to use all available cores in a node due to the limited amount of memory. An additional complication was the need for the user to define a strategy to manually handle the output histogram files and their merging.\\
With the introduction of multi-threading and the improved support of MPI it is now possible to take full advantage of HPC resources: no additional scripting to drive the jobs is required and the output data are naturally merged into a single file. As a demonstrator of the validity of our strategy, we have run test applications on the Babbage test-bed at NERSC.
The results are very promising: we have managed to run an hybrid MPI/MT application on multiple Intel\textsuperscript{\textregistered} Xeon Phi\texttrademark cards, running up to 1000 threads without degradation of performances with respect to a perfect linear speedup. During the tests the only limitation has been the non-optimal use of MPI to merge a large number of histograms. The problem is understood and the final public release Geant4 Version 10.2 will solve this issue.


\section{Conclusions}
Since the introduction of multi-threading, Geant4 has entered the era of massive parallelism: it is the first large HEP code to be migrated to multi-threading. Recent improvements allow for a better integration with external parallelism frameworks and libraries. In particular the Geant4 users have expressed strong interest in TBB library (mainly the HEP community) and for MPI integration (Medical community and HPC users).\\
The integration with these two technologies is substantially improved and is demonstrated in specific examples. In particular an hybrid approach of MPI and MT allows for a simplified use of large core-count resources: the user does not need anymore to write custom scripts to perform job splitting / handling / merging. \\
Preliminary tests performed on the Babbage test-bed at NERSC have shown an excellent scaling of the multi-threading Geant4 scoring with a perfect linearity of the speedup as a function of the number of cores. Further improvements are expected for the MPI merging of results in time for the public release of Geant4 Version 10.2.

\bibliographystyle{IEEEtran}
\bibliography{mybib}{}

\end{document}